\documentclass[12pt,epsf]{article}
\newcommand{\be}{\begin{eqnarray}}
\newcommand{\ee}{\end{eqnarray}}

{
{

\def\0n{0\nu\beta\beta}

\parskip 5pt plus 1pt
\catcode`@=12
\topmargin -0.5in
\evensidemargin 0.0in
\oddsidemargin 0.0in
\textheight 8.5in
\textwidth 6.5in

\begin{document}
\thispagestyle{empty}
\begin{flushright} UCRHEP-T345\\October 2002\
\end{flushright}
\vspace{0.5in}
\begin{center}
{\LARGE \bf Baryon and Lepton Number Violation\\ with Scalar Bilinears \\}
\vspace{1.5in}
{\bf H.~V.~Klapdor-Kleingrothaus$^1$, Ernest Ma$^2$, and Utpal Sarkar$^3$
\\}
\vspace{0.2in}
{$^1$ Max-Planck-Institut f\"ur Kernphysik, D-69029 Heidelberg, Germany} \\ 
\vspace{0.1in}
{$^2$ \sl Physics Department, University of California, Riverside, California 
92521, USA\\}
\vspace{0.1in}
{$^3$ \sl Physical Research Laboratory, Ahmedabad 380 009, India\\}
\vspace{1.5in}
\end{center}

\begin{abstract}

We consider all possible scalar bilinears, which couple to two
fermions of the standard model. The various baryon and lepton
number violating couplings allowed by these exotic scalars
are studied.  We then discuss which ones are constrained by 
limits on proton decay (to a lepton and a meson as well as to 
three leptons), neutron-antineutron oscillations, and neutrinoless 
double beta decay.

\end{abstract}

\vspace{1cm}
\newpage
\baselineskip 24pt
Grand unified theories (GUTs) appear to be the 
most natural extensions of the standard model at the very high
scale. In GUTs the gauge group is unified, so there
is only one single gauge coupling constant to explain all
the forces of the quarks and leptons, which are also treated at the same
footing. However the scale of unification is much too high to 
be directly tested in the laboratory. Thus indirect searches
become very important.  Many predictions of GUTs were
studied, but there is no supporting experimental evidence to date, with the 
possible exception of neutrinoless double beta decay \cite{bb,kk2}. 
These GUT predictions also include new particles such as 
leptoquarks or diquarks and new interactions which violate 
baryon and lepton numbers.  The latter are also required for an 
understanding of the present baryon asymmetry of the Universe.  
An observed Majorana neutrino mass also requires lepton number 
violation.  In the context of these issues we present in this paper 
a general analysis of baryon and lepton number violation with scalar
bilinears. 

The scalar bilinears we consider are generalizations of the usual Higgs 
scalar doublet.  They couple to two fermions of the standard model and 
are present in many of its extensions \cite{2,3,3a}. 
The quantum numbers of these 
particles are thus fixed, but their masses and couplings could be 
arbitrary.  Low-energy particle physics phenomenology (such as the 
non-observation of baryon number violation) could be used 
to constrain their masses and couplings \cite{4,mrs1}.  Baryogenesis 
with scalar bilinears has also been studied \cite{mrs2}. 
Recently a general analysis of the neutrinoless double beta
decay has been performed by constructing effective higher
dimensional operators \cite{chi}. Most of these operators
may be realized in models where scalar bilinears are
included, a few examples of which were demonstrated \cite{chi}.

In Table 1 we list all the scalar bilinears 
allowed in any extension of the standard model. We 
then write down the different baryon and lepton number
violating couplings of these scalars and what kind of
interactions these different couplings may induce, 
leading to a simple classification of all the different 
possible baryon and lepton number violating interactions. 
We include the usual Higgs scalar doublet of the standard
model in these interactions. 

\begin{table}[htb]
\begin{center}
\begin{tabular}{|c|c|c|c|c|c|}
\hline
&&&&&\\
Representation & Notation&
$qq$ & $\bar q \bar l$ & $q \bar l$ & $ll$  \\
&&&&&\\
\hline
$(1,1,-1)$ &$\chi^-$& & & & $\times$ \\
$(1,1,-2)$ &$L^{--}$& & & & $\times$  \\
$(1,3,-1)$ &$\xi$& & & & $\times$ \\
\hline
$(3^*,1,1/3)$ &$Y_a$& $\times$ & $\times$ & &  \\
$(3^*,3,1/3)$ &$Y_b$& $\times$ & $\times$ & &  \\
$(3^*,1,4/3)$ &$Y_c$& $\times$ & $\times$ & &  \\
$(3^*,1,-2/3)$ &$Y_d$& $\times$ & & &  \\
\hline
$(3,2,1/6)$ &$X_a$& & & $\times$ &  \\
$(3,2,7/6)$ &$X_b$& & & $\times$ &  \\
\hline
$(6,1,-2/3)$ &$\Delta_a$& $\times$ & & &  \\
$(6,1,1/3)$ &$\Delta_b$& $\times$ & & &  \\
$(6,1,4/3)$ &$\Delta_c$& $\times$ & & &  \\
\hline
$(6,3,1/3)$ &$\Delta_L$& $\times$ & & &   \\
$(8,2,1/2)$ &$\Sigma$& & & &  \\
\hline
\end{tabular}
\caption{Exotic scalar particles beyond the standard model
and their couplings to combinations of two fermions.}
\end{center}
\label{sc-bil}
\end{table}

For the scalar bilinears, it is not possible to define
definite baryon or lepton number in some cases. However,
depending on their couplings to the combinations of two 
fermions, it is possible to assign baryon and
lepton numbers separately. The different assignments are
given in Table 2.  For any particular 
interaction it will be possible to specify what processes it 
would mediate and what is the change in baryon and lepton
numbers. This will allow us to classify them.

\begin{table}[htb]
\begin{center}
\begin{tabular}{|c|c|c|c|c|}
\hline
&&&&\\
Two-fermion combination &
$qq$ & $\bar q \bar l$ & $q \bar l$ & $ll$  \\
&&&&\\
\hline
&&&&\\
Baryon number& $2/3$ & $-1/3$ & $1/3$ & $ 0$ \\
&&&&\\
Lepton number& $0$ & $-1$& $-1$& $2$ \\
&&&&\\
$B-L$ & $2/3$ & $2/3$ & $4/3$ & $-2$ \\
&&&&\\
\hline
\end{tabular}
\caption{Baryon and lepton numbers of the the scalars coupling
to different combinations of two fermions.}
\end{center}
\label{bil-bl}
\end{table}

The quadratic terms with the scalars are trivial. For any
quadratic term $H_1 H_2$ to be invariant under the gauge
interactions, either both of them are singlets (i.e. neutral under 
all gauge factors) or else $H_1^\dagger = H_2$.  Since all the scalars 
we are discussing are non-singlets, none of the quadratic terms can 
have any $B-L$ quantum number, although some of them may contribute 
to baryon or lepton number violation.  Consider for example,
$Y_a^\dagger Y_a$.  Since $Y_a$ can couple to $\bar q \bar l$
as well as to $q q$, this term may induce a $\Delta B = 1$,
$\Delta L = 1$ process, but $B-L$ remains conserved. 

The trilinear couplings are the most interesting ones. 
It was shown \cite{mrs2} that there are 38 possible 
trilinear couplings with the scalar bilinears. Although
baryon and lepton numbers are not always uniquely specified, all of them
carry a $B-L$ quantum number $\pm 2$. As we shall see next,
these trilinear operators can mediate four possible
processes and hence can be classified in four categories
\begin{itemize}
\item[(i)]  $B=1, L=-1$: In this case both baryon and lepton
numbers are changed by 1, so all these couplings should give
only proton decay, which violates $(B-L)$ \cite{8}. 
\item[(ii)] $B=0, L=2$: This gives lepton number violating
processes, without affecting the baryon number. So this contributes 
to the Majorana masses of the neutrinos. Some of them will 
contribute to the neutrinoless double beta decay \cite{7}, but others
would involve neutrinos of different generations. 
\item[(iii)] $B=1, L=3$: These operators give a three-lepton
decay mode of the proton \cite{8}. In some models these processes
could dominate over the usual proton decay modes.
\item[(iv)] $B=2, L=0$: In this case baryon number is changed
by two unit with no change in lepton number. This corresponds to
neutron-antineutron oscillations.
\end{itemize}
Some of the operators may contribute to two process types. 
There are no trilinear couplings, which conserve $B-L$ (which 
includes of course the case $B=0$ and $L=0$). 

There is only one trilinear coupling which can give proton decay
into a lepton and a meson (or mesons) and is not involved in any 
other process.  It is thus possible to assign $B=1$ and $L=-1$ to 
this operator:

\vbox{$$ \pmatrix{B=1, &  L=-1 } 
$$
\begin{equation}
\begin{array}{l}
{\cal O}_{1} = \mu_{1} X_a \phi Y_d \\
\end{array}
\end{equation}}
In the following $B-L$ quantum numbers for the trilinear couplings are not 
mentioned explicitly since all of them have $B-L=\pm 2$.
All other couplings which contribute to the proton decay,
can also mediate $n - \bar n$ oscillations or the neutrinoless double
beta decay. 

There are only two operators 
which can give rise to the three-lepton decay mode of the proton:

\vbox{$$\pmatrix{B=0,& L=-2 } ~~~\pmatrix{B=1,& L=-1 } 
~~~{\rm and} ~~~ \pmatrix{B=-1,& L=-3 } 
$$
\begin{equation}
\begin{array}{ll}
{\cal O}_2 = \mu_2 Y_a Y_c^\dagger \chi^+ &
{\cal O}_3 = \mu_3 Y_b Y_c^\dagger {\xi}^\dagger \\
\end{array}
\end{equation}}
These operators will simultaneously allow the single-lepton 
decay mode of the proton
($B=1,L=-1$) and neutrinoless double beta decay ($B=0,L=2$). 
There is no trilinear coupling which contributes to the three-lepton
decay mode of the proton and no other process. 

There are four trilinear couplings
which carry $B=2, L=0$ and contribute only to $n - \bar n$
oscillations and not to any other process:

\vbox{$$\pmatrix{B=2,& L=0 } 
$$
\begin{equation}
\begin{array}{ll}
{\cal O}_{4} = \mu_{4} \Delta_a \Delta_b \Delta_b &
{\cal O}_{5} = \mu_{5} \Delta_c \Delta_a \Delta_a \\
{\cal O}_{6} = \mu_{6} \Delta_c Y_d Y_d &
{\cal O}_{7} = \mu_{7} \Delta_L \Delta_L \Delta_a \\
\end{array}
\end{equation}}
Another four trilinear interactions contribute to both $n -
\bar n$ oscillations and proton decay:

\vbox{$$\pmatrix{B=2,& L=0 } 
~~~{\rm and} ~~~ \pmatrix{B=1,& L=-1 } 
$$
\begin{equation}
\begin{array}{ll}
{\cal O}_8 = \mu_8 Y_c Y_d Y_d &
{\cal O}_{9} = \mu_{9} \Delta_L Y_b Y_d \\
{\cal O}_{10} = \mu_{10} \Delta_b Y_a Y_d &
{\cal O}_{11} = \mu_{11} \Delta_a Y_d Y_c \\
\end{array}
\end{equation}}

There are several operators which contribute only
to neutrinoless double beta decay:

\vbox{$$\pmatrix{B=0,& L=2 } $$
\begin{equation}
\begin{array}{lll}
{\cal O}_{12} = \mu_{12} \phi \phi \chi^- & 
{\cal O}_{13} =\mu_{13} \phi \phi \xi &
{\cal O}_{14} = \mu_{14} \chi^- \chi^- L^{++} \\
{\cal O}_{15} = \mu_{15} \xi \xi L^{++}  &
{\cal O}_{16} = \mu_{16} X_b X_a^\dagger \chi^- &
{\cal O}_{17} = \mu_{17} X_b X_a^\dagger \xi \\
{\cal O}_{18} = \mu_{18} \Delta_b X_a^\dagger X_a^\dagger &
{\cal O}_{19} = \mu_{19} \Delta_c X_a^\dagger X_b^\dagger &
{\cal O}_{20} = \mu_{20} \Delta_L X_a^\dagger X_a^\dagger \\
{\cal O}_{21} = \mu_{21} \Delta_a^\dagger \Delta_b \chi^- &
{\cal O}_{22} = \mu_{22} \Delta_a^\dagger \Delta_L \xi &
{\cal O}_{23} = \mu_{23} \Delta_b^\dagger \Delta_c \chi^- \\
{\cal O}_{24} = \mu_{24} \Delta_L^\dagger \Delta_c \xi &
{\cal O}_{25} = \mu_{25} \Delta_a^\dagger \Delta_c L^{--} & \\
\end{array}
\end{equation}}
There are few more operators contributing to neutrinoless double
beta decay, which also allow proton decay:

\vbox{$$\pmatrix{B=0,& L=2 } 
~~~{\rm and} ~~~ \pmatrix{B=-1,& L=1 } 
$$
\begin{equation}
\begin{array}{lll}
{\cal O}_{26} = \mu_{26} Y_a Y_d^\dagger {\chi^-} &
{\cal O}_{27} = \mu_{27} Y_b Y_d^\dagger {\xi} &
{\cal O}_{28} = \mu_{28} Y_c Y_d^\dagger L^{--}  \\
{\cal O}_{29} = \mu_{29} X_a X_b Y_c^\dagger &
{\cal O}_{30} = \mu_{30} X_a \phi^\dagger Y_a &
{\cal O}_{31} = \mu_{31} X_a \phi^\dagger Y_b \\
{\cal O}_{32} = \mu_{32} X_a X_a Y_a^\dagger &
{\cal O}_{33} = \mu_{33} X_a X_a Y_b^\dagger &
{\cal O}_{34} = \mu_{34} X_b Y_d \phi^\dagger \\
\end{array}
\end{equation}}
The proton decay
constraints would then restrict the couplings of these operators 
making them very much suppressed. However, if some of the couplings could be
avoided with discrete symmetries, then these operators could also 
contribute to the neutrinoless double beta decay significantly.
Finally there are four more operators which contribute to 
neutrinoless double beta decay, proton decay, and $n - \bar n$
oscillations simultaneously:

\vbox{$$\pmatrix{B=0,& L=-2 }, ~~~ \pmatrix{B=1,& L=-1 }
~~~{\rm and} ~~~ \pmatrix{B=2,& L=0 } 
$$
\begin{equation}
\begin{array}{ll}
{\cal O}_{35} = \mu_{35} Y_d Y_a Y_a &
{\cal O}_{36} = \mu_{36} Y_d Y_b Y_b \\ 
{\cal O}_{37} = \mu_{37} \Delta_a Y_a Y_a &
{\cal O}_{38} = \mu_{38} \Delta_a Y_b Y_b \\
\end{array}
\end{equation}}

The quartic terms belong to two broad classes, one with $B-L=0$ and
the other with $B-L=4$. Again they can be categorized under certain 
subclasses.  Those which are products of the usual 
quadratic couplings of the form $\Phi^\dagger \Phi$ belong to the
trivial category with $B=0$, $L=0$ (and thus $B-L=0$), which
we shall not list here.  All the quartic couplings 
involving only the dileptons ($\chi^-, L^{--}$, and $\xi$), the 
usual Higgs doublet $\phi$, and the octet $\Sigma$ also fall 
into this trivial category with $B=0$, $L=0$. 
There are also some other quartic couplings involving the diquarks and
leptoquarks, which can be classified under this trivial
category. These are 
$$B-L=0 ~~~~\pmatrix{B=0,& L=0} $$
\begin{equation}
\begin{array}{lll}
\widetilde{\cal O}_1 = \chi^+ \chi^+ \xi \xi & 
\widetilde{\cal O}_2 = \phi \phi \chi^+ L^{--}
& \widetilde{\cal O}_3 = \Sigma \Sigma \chi^+ L^{--} \\
\widetilde{\cal O}_4 = \phi^\dagger \phi^\dagger \xi L^{++} & 
\widetilde{\cal O}_5 = \Sigma \Sigma \xi^\dagger L^{--} & 
\widetilde{\cal O}_6 = \phi^\dagger \phi^\dagger \Sigma \Sigma \\
\widetilde{\cal O}_7 = \Delta_a^\dagger X_a X_a \chi^- & 
\widetilde{\cal O}_8 = \Delta_a^\dagger X_a X_a
\xi & \widetilde{\cal O}_9 = \Delta_b^\dagger X_b X_b L^{--} \\
\widetilde{\cal O}_{10} = \Delta_c^\dagger X_b X_b \chi^- & 
\widetilde{\cal O}_{11} = \Delta_c^\dagger X_b X_b
\xi & \widetilde{\cal O}_{12} = \Delta_L^\dagger X_b X_b L^{--} \\
&\widetilde{\cal O}_{13} = \Delta_a^\dagger X_a X_b L^{--} &\\ 
\end{array}
\end{equation}

There are several $B-L=0$ quartic couplings which 
mediate baryon number violating processes. The simplest of them have 
$B=2$ and $L=2$.  The
hydrogen-antihydrogen ($p+e^- \to \bar p + e^+$)
oscillations and double proton decay into two
positrons ($p + p \to e^+ + e^+$) are typical examples of such
processes. There are 7 such interactions
$$ B-L=0 ~~~~ \pmatrix{B=2,& L=2}$$ 
\begin{equation}
\begin{array}{lll}
\widetilde{\cal O}_{14} = \Delta_a \Delta_c \Delta_c L^{--} & 
\widetilde{\cal O}_{15} = \Delta_b \Delta_b 
\Delta_b \chi^- & 
\widetilde{\cal O}_{16} = \Delta_b \Delta_b \Delta_L \xi \\
\widetilde{\cal O}_{17} = \Delta_b \Delta_L \Delta_L \chi^- & 
\widetilde{\cal O}_{18} = \Delta_b \Delta_L
\Delta_L \xi & \widetilde{\cal O}_{19} = \Delta_L \Delta_L \Delta_L \chi^- \\
& \widetilde{\cal O}_{20} = \Delta_L \Delta_L \Delta_L \xi &\\
\end{array}
\end{equation}
There are other quartic couplings also, which allow these 
processes with $B=2$ and $L=2$. They involve $Y_a$,
$Y_b$, and $Y_c$, which couple to two quarks as well as to an antiquark 
and an antilepton. Hence some of these interactions can simultaneously allow 
$B=1$ and $L=1$ 
proton decay processes such as $p \to e^+$ or $n \to \bar \nu$
associated with a neutral meson and a lepton-antilepton pair.
In all these interactions, the constraints from proton decay will not 
allow processes with $B=2$.  These operators are 
$$ B-L=0 ~~ \pmatrix{B=2,& L=2 } ~~~ {\rm and} ~~~ 
\pmatrix{B=1,& L=1} $$
\begin{equation}
\begin{array}{ll}
\widetilde{\cal O}_{21} = \Delta_c Y_a Y_d \chi^- & 
\widetilde{\cal O}_{22} = \Delta_c Y_b Y_d \xi \\
\end{array}
\end{equation}
\vskip.1in
\vbox{
$$ B-L=0 ~~ \pmatrix{B=2 \cr L=2 }, ~~~
\pmatrix{B=1 \cr L=1 } ~~~ {\rm and} ~~~ 
\pmatrix{B=0 \cr L=0 }
$$
\begin{equation}
\begin{array}{lll}
\widetilde{\cal O}_{23} = \Delta_a Y_a Y_c \chi^- & 
\widetilde{\cal O}_{24} = \Delta_a Y_b Y_c \xi & 
\widetilde{\cal O}_{25} = \Delta_b Y_a Y_c L^{--} \\ 
\widetilde{\cal O}_{26} = \Delta_b Y_a Y_a \chi^{-} &
\widetilde{\cal O}_{27} = \Delta_b Y_a Y_b \xi& 
\widetilde{\cal O}_{28} = \Delta_b Y_b Y_b \chi^{-} \\ 
\widetilde{\cal O}_{29} = \Delta_b Y_b Y_b \xi & 
\widetilde{\cal O}_{30} = \Delta_c Y_a Y_a L^{--} &
\widetilde{\cal O}_{31} = \Delta_c Y_a Y_a L^{--} \\ 
\widetilde{\cal O}_{32} = \Delta_L Y_a Y_a \xi &
\widetilde{\cal O}_{33} = \Delta_L Y_a Y_b \chi^- & 
\widetilde{\cal O}_{34} = \Delta_L Y_a Y_b \xi \\
\widetilde{\cal O}_{35} =  \Delta_L Y_b Y_b \chi^- & 
\widetilde{\cal O}_{36} = \Delta_L Y_b Y_b \xi &
\widetilde{\cal O}_{37} = \Delta_L Y_b Y_c  L^{--} \\ 
\widetilde{\cal O}_{38} = Y_a Y_c Y_d \chi^- &
\widetilde{\cal O}_{39} = Y_b Y_c Y_d \xi & 
\widetilde{\cal O}_{40} = Y_c Y_d Y_d L^{--} \\
\end{array}
\end{equation}}
\vskip .1in
\vbox{
$$ 
B-L=0 ~~ \pmatrix{B=2 \cr L=2 }, ~
\pmatrix{B=1 \cr L=1 }, ~ 
\pmatrix{B=0 \cr L=0 } ~ {\rm and} ~ 
\pmatrix{B=-1 \cr L=-1}
$$
\begin{equation}
\begin{array}{lll}
\widetilde{\cal O}_{41} = Y_a Y_a Y_a \chi^- & 
\widetilde{\cal O}_{42} = Y_a Y_a Y_b \xi & 
\widetilde{\cal O}_{43} = Y_a Y_a Y_c L^{--} \\
\widetilde{\cal O}_{44} = Y_a Y_b Y_b \chi^- & 
\widetilde{\cal O}_{45} = Y_a Y_b Y_b \xi & 
\widetilde{\cal O}_{46} = Y_b Y_b Y_b \chi^- \\ 
\widetilde{\cal O}_{47} = Y_b Y_b Y_b \xi & 
\widetilde{\cal O}_{48} = Y_b Y_b Y_c L^{--} &\\ 
\end{array}
\end{equation}}

Another class of quartic couplings can give rise to $B-L=4$
processes. Some of them can produce only $B=2$ and $L=-2$ processes
like $n + n \to e^- e^- \pi^+ \pi^+$ or $n - \bar n$ oscillation
associated with charged leptons and charged pions. The couplings which 
can give rise to these processes are

\vbox{
$$ B-L=4 ~~~~ \pmatrix{B=2,& L=-2} $$ 
\begin{equation}
\begin{array}{ll}
\widehat{\cal O}_{1} = Y_d Y_d Y_d L^{++} & 
\widehat{\cal O}_2 = \Delta_a Y_d Y_d L^{++} \\
\widehat{\cal O}_3 = \Delta_a \Delta_a \Delta_a L^{++} & 
\widehat{\cal O}_4 = \Delta_a \Delta_a \Delta_b \chi^+ \\
\end{array}
\end{equation}}
The other subclass of $B-L=4$ processes have $B=1$ and $L=-3$, which
give rise to $p \to \nu \nu \nu \pi^+$ for example.  
There are two such operators,

\vbox{$$ B-L=4 ~~~~ \pmatrix{B=1,& L=-3} $$ 
\begin{equation}
\begin{array}{ll}
\widehat{\cal O}_5 = X_a X_a X_a \phi^\dagger & 
\widehat{\cal O}_6 = X_a X_a X_a \Sigma^\dagger \\
\end{array}
\end{equation}}
There are also some quartic couplings with $B-L=4$, which allow both
of the above two interactions. There are three such terms 

\vbox{ $$ B-L=4 ~~~~ \pmatrix{B=1,& L=-3 } ~~~ {\rm and} ~~~
\pmatrix{B=2,& L=-2 }  $$
\begin{equation}
\begin{array}{lll}
\widehat{\cal O}_7 = Y_a Y_d Y_d \chi^+ & 
\widehat{\cal O}_8 = \Delta_a Y_a Y_d \chi^+ & 
\widehat{\cal O}_9 = 
\Delta_a Y_b Y_d \xi^\dagger \\
\end{array}
\end{equation}}

In summary, we have made a general model-independent study of the 
possible baryon and lepton number violations beyond the standard model 
with scalar bilinears.  Depending on the processes under consideration, 
they can be classified into only a few categories.  We list all the
trilinear and quartic couplings and discuss the processes to which    
these operators contribute. 

\vskip .2in
\centerline{\bf Acknowledgement}

\vspace{0.5cm}
\noindent
Two of us (EM and US) thank the Max-Planck-Institut f\"ur Kernphysik 
for hospitality. The work of EM is supported in part by the 
U.S. Department of Energy under Grant No. DE-FG03-94ER40837.

\end{document}